\documentclass[aps,superscriptaddress,floatfix,showpacs]{revtex4}
\usepackage{graphicx}% Include figure files
\usepackage[english]{babel}
\usepackage[usenames]{color}
\usepackage{amsmath,amsfonts,amssymb,float}
\usepackage{dcolumn}% Align table columns on decimal point
\usepackage{bm}% bold math

\begin{document}

\title{Efficient Raman amplification into the PetaWatt regime}

\author{R.M.G.M. Trines}
\affiliation{Rutherford Appleton Laboratory, Harwell Science and Innovation Campus, Didcot, Oxon, OX11 0QX, United Kingdom}
\author{F. Fi\'uza}
\affiliation{GoLP/Instituto de Plasmas e Fus\~ao Nuclear, Instituto Superior
  T\'ecnico, 1049-001 Lisbon, Portugal}
\author{R. Bingham}
\altaffiliation{also at University of Strathclyde, Glasgow, G4 0NG,
 United Kingdom}
\affiliation{Rutherford Appleton Laboratory, Harwell Science and Innovation Campus, Didcot, Oxon, OX11 0QX, United Kingdom}
\author{R.A. Fonseca}
\author{L.O. Silva}
\affiliation{GoLP/Instituto de Plasmas e Fus\~ao Nuclear, Instituto Superior
  T\'ecnico, 1049-001 Lisbon, Portugal}
\author{R.A. Cairns}
\affiliation{University of St Andrews, St Andrews, Fife KY16 9AJ,
 United Kingdom}
\author{P.A. Norreys}
\affiliation{Rutherford Appleton Laboratory, Harwell Science and Innovation Campus, Didcot, Oxon, OX11 0QX, United Kingdom}
\date\today

\begin{abstract}

  Raman amplification of a short laser pulse off a long laser beam has been
  demonstrated successfully for moderate probe intensities ($\sim 10^{16}$
  W/cm$^2$) and widths ($\sim 50$ micron). However, truly competitive
  intensities can only be reached if the amplification process is carried out
  at much higher probe intensities ($10^{17}-10^{18}$ W/cm$^2$ after
  amplification) and widths ($1-10$ mm). We examine the opportunities and
  challenges provided by this regime through the first 2-dimensional
  particle-in-cell simulations using wide pulses. A parameter window is
  identified in which a 10 TW, 600 $\mu$m wide, 25 ps long laser pulse can be
  efficiently amplified to 2 PW peak intensity.

\end{abstract}

\maketitle

\def\pd#1#2{\frac{\partial #1}{\partial #2}}
\def\pdd#1#2{\frac{\partial^2 #1}{\partial #2 ^2}}
\def\D{\mathrm{d}}
\renewcommand{\vec}[1]{\mathbf{#1}}

There has been a great deal of interest in obtaining ultra-high laser
intensities by using the technique of Raman amplification in a plasma
\cite{shvets98,shvets99}. The main reason for using a plasma is that it can
tolerate much higher laser intensities ($10^{17}$ W/cm$^2$ or more) than solid
state devices that are commonly used in the Chirped Pulse Amplification (CPA)
scheme (up to $10^{12}$ W/cm$^2$). Extensive analytical
\cite{shvets98,shvets99,malkin00,malk00pop,tsid00,tsid02,balakin03,ersfeld},
numerical \cite{leehj,mardahl,mshur04,clark05,balakin05,mshur05,weber} and
experimental \cite{ping04,dreher04,cheng05,kirkw07} research recently
culminated in the experimental demonstration of Raman amplification of a of a
$10^{12}$ W/cm$^2$ probe pulse to $10^{16}$ W/cm$^2$ off a $10^{14}$ W/cm$^2$
pump pulse in a 1 mm long, 80 $\mu$m wide plasma channel \cite{ren07}. Based
on these results, it has been predicted that the Raman amplification scheme
can be extended to produce laser pulses at much higher intensities than
currently available from solid-state lasers.  Eventually, this should lead to
Raman amplification of cm-wide laser beams having $\sim 1$ $\mu$m wavelength
to $10^{17}$-$10^{18}$ W/cm$^2$ \cite{malkin00}, which are to be focused down
to $\mu$m-wide spots to reach peak intensities of $10^{25}$-$10^{29}$ W/cm$^2$
\cite{fisch03,malkin05,malkin07}, approaching the Schwinger limit.

In order to reach such extreme intensities, the Raman amplification scheme
needs to be extended from its current dimensions (1 mm long, 50-80 $\mu$m
wide, $10^{14}$ W/cm$^2$ pump, $10^{16}$ W/cm$^2$ final probe) to a length and
width of 1-10 cm each, a pump intensity of $10^{14}$-$10^{16}$ W/cm$^2$ and a
final probe intensity of $10^{17}$-$10^{18}$ W/cm$^2$. Quite significant
extensions of the scheme are necessary in any event, to investigate whether
Raman amplification will be able to compete with recent developments in
solid-state laser technology, e.g. OPCPA \cite{ross97,chekhlov07}. However,
there are few results in the literature on the feasibility of such extensions
\cite{shvets99,clark05}. The existing analytic theory is one-dimensional and
only weakly non-linear \cite{shvets98,shvets99,malkin00}. Most numerical
simulations have been performed in one dimension and with envelope models for
the laser pulses \cite{mshur04,balakin05,mshur05} rather than solving the full
set of Maxwell's equations \cite{leehj,clark05,weber}. Only a few 2-D
particle-in-cell (PIC) simulation results are available, and these only deal
with narrow pulses (4-6 $\mu$m) and short propagation distances ($\simeq 1$
mm) \cite{mardahl}.  Experiments whose parameters go beyond that of Ren
\emph{et al.} \cite{ren07} have yet to be conducted. In short, while there is
a strong need for any results concerning the extension of Raman amplification
to greater propagation lengths, pulse widths and pulse intensities, such
results have not been available until now.

\begin{figure}
\includegraphics[width=0.5\textwidth]{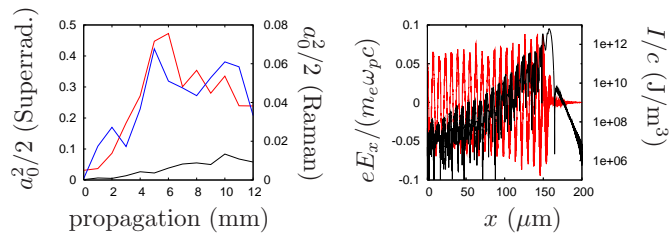}
\caption{Effects of increasing the interaction length. Left: Probe intensity
  $a_0^2/2$ versus propagation distance in mm for Raman amplification with
  $a_0 = a_1 = 0.1$ and $\omega_0/\omega_p = 10$ (red curve, right y-axis) and
  the two superradiant scenarios of Ref. \cite{shvets98} (black and blue
  curves, left y-axis). Both the Raman and the high-density superradiant
  scenario (blue) saturate after about 5 mm, while probe growth is too slow to
  reach saturation for the low-density superradiant scenario (black).  Right:
  RFS and modulational instability for Raman amplification with $a_0 = 0.01$,
  $a_1 = 0.1$ and $\omega_0/\omega_p = 10$. The probe loses energy by driving
  a wakefield, visible in the longitudinal electric field (red), while its
  intensity envelope (black) is modulated following the wakefield period.}
\label{fig:1}
\end{figure}

In this paper we present the results of a series of
multi-dimensional particle-in-cell simulations that have been conducted to
investigate the possibilities for extending the Raman amplification scheme.
These simulations have been carried out at greater propagation lengths, pulse
widths and pulse intensities than have been studied in previous publications,
to investigate the effects of these on the amplification process. A multitude
of nonlinear effects have been encountered, such as probe saturation due to
Raman forward scattering (RFS) and wakefield generation, breaking of the RBS
Langmuir wave that couples pump and probe, parasitic pump RBS, and transverse
filamentation of both pump and probe pulses in 2-D simulations. For each
simulation we have also determined the efficiency of the energy transfer from
pump to probe, and found that this efficiency is critically dependent on a
number of parameters. Controlling the efficiency is vital for the success of
the scheme, as most predictions regarding extreme probe intensities are based
on the assumption of near-perfect efficiency.  Although these issues narrow
the parameter window for effective Raman amplification down considerably, we
have been able to identify a parameter regime in which a 10 TW, 600 $\mu$m
wide, 25 ps long laser pulse can be efficiently amplified to 2 PW peak
intensity, as discussed below.

For the simulations, we have used the 1-D and 2-D versions of the
particle-in-cell (PIC) codes XOOPIC \cite{oopic} and OSIRIS 2.0 \cite{osiris}.
XOOPIC has been used to study Raman amplification before \cite{leehj,mardahl}
and can launch a backward-moving long pump pulse from the leading edge of a
moving window. This allowed us to use a small moving window following the
probe pulse while it interacts with a long pump, thus concentrating on the
evolution of the probe. On the other hand, OSIRIS is fully parallelized, so it
could be used to study the entire evolution of the pump pulse during its
propagation through the plasma column, before meeting the probe pulse. Where
possible, we have used the simulation results of both codes for mutual
verification, thus ensuring their correctness.

\begin{figure}
\includegraphics[width=0.5\textwidth]{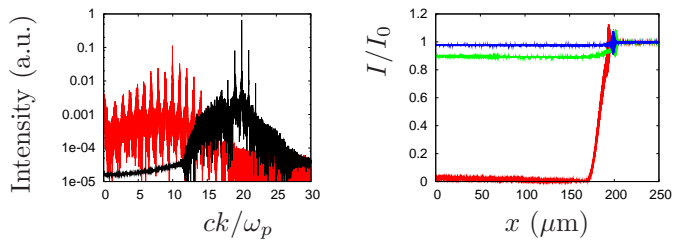}
\caption{(a) Wave number spectrum of the pump pulse in a 5.4 mm plasma with
  $a_0 = 0.1$ and $\omega_0/\omega_p = 10$ (red) or 20 (black). At the higher
  plasma density, the presence of a large number of (anti-)Stokes satellites
  indicates strong parasitic SRS and pump modulation, rendering the pump
  unsuitable for amplification. At the lower density, these effects are much
  less pronounced and do not unduly affect the amplification process. (b)
  Effect of decreasing plasma density on energy transfer efficiency. Shown is
  the depletion of the pump for $\omega_0/\omega_p = 10$ (red), 20 (green) and
  40 (blue).}
\label{fig:2}
\end{figure}

To investigate the effect of a longer interaction length, we have performed a
series of 1-D simulations using both XOOPIC and OSIRIS. Parameters of these
simulations were as follows. Pump and probe amplitudes were $a_0 = a_1 = 0.1$,
pump wave length $\lambda_0 = 800$ nm, $\omega_0/\omega_p = 10$, the pump was
``infinitely'' long while the initial probe duration was 50 fs. The results
are depicted in Figure \ref{fig:1}. The left frame shows the evolution of the
probe intensity for several typical amplification scenarios. It was found that
the probe amplification starts off in a promising way, but usually saturates
after up to 5 mm of propagation, when its intensity becomes sufficient to
trigger nonlinear effects (red and blue curves). Saturation was only absent
when the energy transfer was so inefficient that the probe was not properly
amplified (black curve). Saturation happens for the following reasons: probe
RFS (which scatters energy away from the matching probe frequency to
non-matching frequencies that cannot contribute to the Raman amplification
process), modulational instability of the probe and subsequent wakefield
generation by the probe (which combine to deplete the energy of the probe).
The effects of RFS and probe modulation on Raman amplification are displayed
in the right frame of Figure \ref{fig:1}. It is found that the probe leaves a
wakefield behind, while its envelope is modulated on the wave length of the
wakefield. RFS and probe modulation therefore need to be avoided at all cost.
This can be done by decreasing the plasma density so the probe length is
shorter than $2\pi c/\omega_p$ and RFS growth is inhibited, as discussed
below. There is no real cure for the saturation of the probe; it simply limits
the intensity to which the probe can be Raman amplified, as already predicted
in Ref.  \cite{malk00pop}. In our simulations, we found that efficient probe
amplification is possible until the probe intensity is about 200 times that of
the pump. Beyond that, saturation sets in, the efficiency of the process drops
and the probe envelope deteriorates.

In addition, it has been found that the pump may suffer from parasitic
instabilities as it penetrates the plasma column even before meeting the
probe, mostly parasitic RBS, as illustrated in Figure \ref{fig:2}, left. Here,
the wave number spectra of pump pulses having $a_0 = 0.1$ are shown after
traversing a 5 mm long plasma column at $\omega_0/\omega_p = 10$, 20. At the
higher density, the pump's spectrum shows many side bands which are not much
below the intensity of the fundamental peak. This renders the pump useless for
Raman amplification. At the lower density, the pump's spectrum shows fewer
side bands at lower relative intensity, indicating that the pump is still
suitable for amplification in this case. Thus, lowering the plasma density
will mitigate both probe and pump instabilities and is therefore recommended.
However, in simulations using $a_0 = 0.01$, $a_1 = 0.1$, the efficiency of the
amplification process was found to be 90-100\%, 5\%, 1\% for
$\omega_0/\omega_p = 10$, 20, 40 respectively, as shown in Figure \ref{fig:2},
right. Similar results were recovered from simulating the two scenarios for
``superradiant'' amplification from Ref. \cite{shvets98} (see Figure
\ref{fig:1}, left): in the high-density scenario the probe pulse grows quickly
and saturates after several mm, while the low-density scenario is
characterised by poor energy transfer efficiency and slow probe growth and
saturation is never reached. This will be shown to be a recurrent theme:
measures that reduce the impact of ``bad'' instabilities often also reduce the
effectiveness of the ``good'' instabilities, rendering it difficult (although
not impossible) to strike the right balance between the two. As will be
discussed below, the efficiency problem can be reduced by increasing the
intensities of pump and probe, under specific conditions.

Chirping the pump pulse to induce a frequency mismatch that suppresses
parasitic RBS \cite{malkin00} is not necessarily practical: maintaining a
decent frequency gradient over a large pump length requires a fairly large
bandwidth.  Compressing such a pulse can easily be done using a grating with a
large surface, thus bypassing the more complex Raman amplification process
altogether.

\begin{figure}
\includegraphics[width=0.5\textwidth]{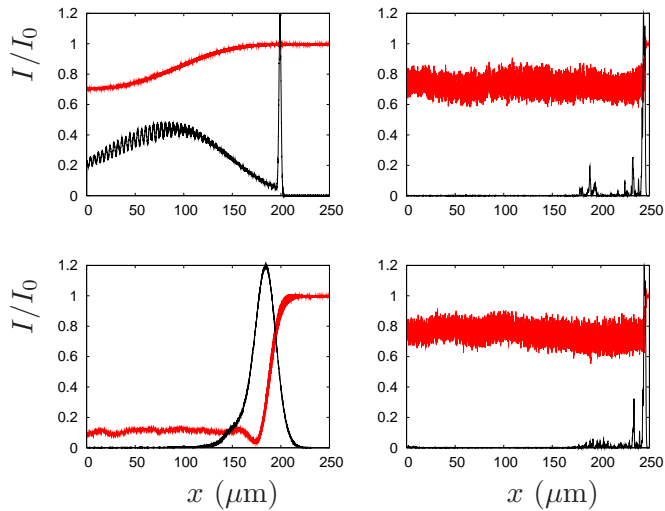}
\caption{Effects of pump intensity of growth of long and short probes. Shown
  are the relative intensities of pump (red) and probe (black) versus
  longitudinal coordinate in metres. Top row: 50 fs probe and pump with $a_0 =
  0.01$, 0.1 respectively. The less intense pump causes the short probe to
  stretch, while the more intense pump causes the probe to remain short.
  Bottom row: Same as before, but with 500 fs probe. At low pump intensity the
  probe remains long, while at higher pump intensity the probe is shortened
  due to breaking of the RBS Langmuir wave that couples pump and probe. Best
  efficiency is obtained for long probe and low-intensity pump; efficiency is
  much lower for more intense pumps.}
\label{fig:3}
\end{figure}

The intensity of the pump affects the amplification process in many ways. The
use of lower pump intensities ($\leq 10^{14}$ W/cm$^2$) will of course reduce
the growth rate of parasitic instabilities. However, a longer pump and thicker
plasma slab will be needed to amplify the probe to a similar final intensity,
which allows parasitic instabilities more time to grow. On balance, we found
that it is often better to increase the pump intensity and reduce the
thickness of the plasma slab, as this leads to a smaller total influence of
parasitic instabilities. In addition, lower pump intensities favour the
amplification of longer probes, while increasing the pump intensity will lead
to a shorter probe after amplification. Even if a short initial probe is used,
the narrowband linear amplification that occurs at low intensities will reduce
the probe's bandwidth and increase its length, as predicted in Ref.
\cite{shvets99}. For higher pump intensities ($\gtrsim 10^{15}$ W/cm$^2$), the
amplification process will be much faster, so the pump depletion length will
be shorter and the amplification of shorter probes will be favoured. However,
when the pump is too intense ($\gtrsim 10^{16}$ W/cm$^2$) the amplification of
the probe may be curbed by wave breaking of the RBS Langmuir wave. This will
happen when the combined pump and probe intensities become too large, and will
stop the energy flow from pump to probe long before the pump is fully
depleted, thus sharply reducing the efficiency of the process. The effect of
the pump intensity on the length of the amplified probe is illustrated by the
simulation results depicted in Figure \ref{fig:3}.

\begin{figure}
\includegraphics[width=0.5\textwidth]{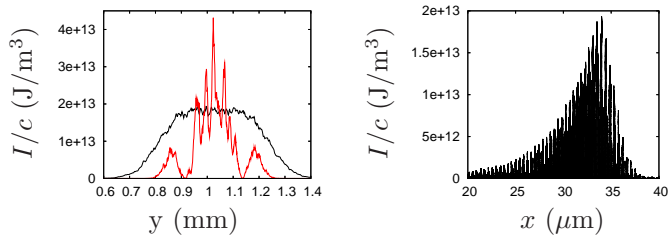}
\caption{Left: effect of filamentation on transverse probe profile for
  $\omega_0/\omega_p = 10$ and 300 $\mu$m probe width after 2.4 mm (red) and
  $\omega_0/\omega_p = 20$ and 600 $\mu$m probe width after 4 mm (black). In
  both cases, $a_0 = 0.03$ and $a_1 = 0.1$. The runaway
  filamentation at high plasma density renders the probe useless, while the
  temperate filamentation at lower density does not unduly compromise probe
  focusability. Right: longitudinal probe profile from the simulation at
  $\omega_0/\omega_p = 20$, showing that a peak intensity of over $3\times
  10^{17}$ W/cm$^2$ can be reached in 25-30 fs.}
\label{fig:4}
\end{figure}

Since the maximum amplification/compression factor is limited to about 200 by
saturation, while the pump intensity cannot be too high because of pump
instabilities and Langmuir wave breaking, the most reliable way to increase
the total energy of the amplified probe is to attempt to amplify wider pulses.
This requires a thorough understanding of all transverse effects that occur
during Raman amplification. Until now, transverse effects in Raman
amplification have not received much attention in theoretical studies. Some
3-D hydrodynamic simulations have been conducted \cite{balakin03,balakin05},
but these do not include the full gamut of transverse laser-plasma
instabilities. The only available 2-D PIC simulations \cite{mardahl} restrict
themselves to narrow pulses (4-6 $\mu$m) and short propagation distances
($\simeq 1$ mm) in a preformed plasma channel. However, much wider pulses are
needed (1-10 mm and beyond) to push Raman amplification to truly high
intensities \cite{fisch03,malkin05,malkin07}. Pulses of such width may be
affected by various transverse phenomena, such as self-focusing
\cite{sprangle87}, filamentation \cite{kaw73} and gain narrowing (when the
centre of the probe is amplified more efficiently than the wings because of
its intrinsically higher intensity, and an effective narrowing of the probe
occurs). To study these effects we have carried out the first 2-D full-PIC
simulations ever that use wide pulses (widths of 300 and 600 $\mu$m) in a
homogeneous plasma. The results of these simulations are displayed in Figure
\ref{fig:4}, which shows the amplified probe intensity profiles resulting from
these simulations. Here, the pump intensity was $10^{15}$ W/cm$^2$, the probe
intensity was $10^{16}$ W/cm$^2$, the probe duration was 50 fs and
$\omega_0/\omega_p = 10$ or 20 was used, with a flat transverse density
profile. For both densities, the probe was amplified to several times
$10^{17}$ W/cm$^2$, but it also suffered from transverse filamentation, in
particular for $\omega_0/\omega_p = 10$. During early stages of the
amplification, a transverse envelope modulation with a period of $2\pi
c/\omega_p$ emerges, while merging of narrow filaments into wider ones occurs
during later stages.  For $\omega_0/\omega_p = 10$, this effect is so strong
that complete filamentation occurs after only 2.4 mm propagation. For
$\omega_0/\omega_p = 20$ the effect is much less prominent, which allowed the
probe to reach $\sim 5\times 10^{17}$ W/cm$^2$ in 25-30 fs after 4 mm of
propagation before its envelope started to break up. For a 600 $\mu$m wide
probe, this implies a peak power of around 2 PetaWatt, with an estimated
efficiency of 30-40 \%. Note that the probe growth would have saturated for
longer propagation distances anyway, so transverse effects need not limit the
maximum amplification even further for the right set of parameters. This
particular scenario has several characteristics that make it work. The plasma
density is fairly low, so the probe fits into a single plasma period,
inhibiting RFS and modulational instabilities. In addition, pump and probe
instabilities do not grow out of control for several mm of propagation. Pump
and probe intensities are fairly high, so the probe is rapidly and efficiently
amplified over a short distance, but they do not yet trigger severe growth of
parasitic instabilities or Langmuir wave breaking at this density. Simulations
of a more intense pump ($10^{16}$ W/cm$^2$) revealed filamentation of the pump
front after 2 mm, before meeting the probe, emphasizing the importance of
using the right combination of pump intensity, plasma density and propagation
distance. Most importantly, there is no fundamental disadvantage to extending
this scenario to pulses much wider than 600 $\mu$m.

Although the results at $\omega_0/\omega_p = 10$ indicate that this density is
too high for the amplification of wide pulses, they still provide important
information on the transverse behaviour of the amplified probe. Merging of
filaments and self-focusing of individual filaments can be observed, but there
is no evidence of whole-beam self-focusing (outlying filaments are not pushed
towards the pulse centre during the first few mm of propagation). This is in
line with the observation that whole-beam relativistic self-focusing is
inhibited by the transverse density modulation that accompanies the
filamentation instability. There is no evidence of gain narrowing: the
relative amplification of individual filaments does not appear to depend on
their transverse position. Based on these results, we predict that the
narrowing of the probe pulse observed by Ren \emph{et al.} \cite{ren07} stems
from focusing of the probe in the preformed plasma channel rather than gain
narrowing or relativistic self-focusing, as the amplified probe is
sufficiently narrow with respect to $2\pi c/\omega_p$ to behave as a single
filament.

Pulse filamentation puts restrictions on the range of useful pump intensities:
too low, and the probe will have to propagate in plasma for too long, causing
it to filament; too high, and the pump will filament itself. Probe
filamentation will also limit the extent to which the probe can be focused, so
the probe amplification has to be stopped before it grows too large.
Nevertheless, good amplification can still be obtained with a proper choice of
parameters. For example, the results displayed in Figure \ref{fig:4} show that
a 10 TW, 20 ps pump of up to 1 mm diameter, containing 200 J, can be
compressed to less than 100 fs and up to 2 PW using a homogeneous plasma
column of several mm long and wide, which appears experimentally feasible. The
maximum amplification of the probe is limited by nonlinear saturation and
again by filamentation; realistically, the probe can be amplified to about 200
times the intensity of the pump before either effect sets in. Therefore, the
only way to increase the energy content of the amplified probe is to increase
the transverse dimensions of pulses and plasma while keeping the plasma slab
thin. Although it is not possible to do this for both transverse directions,
it may be achieved for one direction using an elongated supersonic gas flow
and a line focus for the pulses.

In this paper, we have shown for the first time, through numerical
simulations, that there are limits to Raman amplification set by various
instabilities of both pump and probe pulses. This reduces the useful ranges of
the experimental parameters. If the plasma density is too low
($\omega/\omega_p \sim 40$), the energy transfer is inefficient; too high
($\omega/\omega_p \sim 10$), and filamentation will destroy the pulse
profiles. If the pump intensity is too low ($\sim 10^{14}$ W/cm$^2$), the
probe will take too long to amplify, allowing probe filamentation and RFS to
grow out of control; too high ($\sim 10^{16}$ W/cm$^2$), and the pump will be
wasted by parasitic RBS and filamentation even before it meets the probe. The
length of the plasma column is limited to a few mm by probe saturation and
filamentation and various pump instabilities. For an ideal choice of
parameters, probe saturation and filamentation would occur after roughly the
same propagation distance, while both the plasma density and pump intensity
would be high enough to allow for efficient amplification, low enough to
inhibit pump instabilities occurring over the length of the plasma column. We
have found that probe amplification up to the PetaWatt level is possible in 4
mm for $\omega/\omega_p = 20$, a $\sim 10^{15}$ W/cm$^2$ pump, and 600 $\mu$m
wide pulses. Even so, low-level filamentation will limit the focusability and
thus the peak intensity of the probe. Since the proposed scenarios for Raman
amplification beyond $10^{25}$ W/cm$^2$ rely on either extreme focusability
\cite{fisch03,malkin07}, very long propagation lengths and intense pumps
\cite{malkin05}, or a very high plasma density ($\omega_0/\omega_p < 5$)
\cite{malkin07}, they are not feasible in light of our new simulation results,
and the Schwinger limit is out of reach for the foreseeable future. However,
if the limits to Raman amplification set by the various instabilities are
properly observed, there is a narrow but definite parameter window in which
good amplification can be obtained.

This work was supported by the STFC Accelerator Science and Technology Centre
and the STFC Centre for Fundamental physics, and by FCT (Portugal) through
grant PTDC/FIS/66823/2006. We would like to thank W. Mori and D. Jaroszynski
for useful discussions, the Plasma Theory and Simulation Group of UC Berkeley
for the use of XOOPIC, and the OSIRIS consortium for the use of OSIRIS. Some
of the simulations were performed using the IST Cluster, Lisbon.

\end{document}